\documentclass{ws-procs975x65}
\usepackage{amsmath}
\usepackage{amsfonts}
\usepackage{amssymb}
\usepackage[dvips]{epsfig}

\begin{document}

\title{A Note on the Robustness of Pair Separations Methods
in Cosmic Topology}

\author{A. Bernui, \ \ B. Mota, \ \  M.J. Rebou\c{c}as}
\address{Centro Brasileiro de Pesquisas F\'{\i}sicas\\
Rua Dr.\ Xavier Sigaud 150, \ \  22290-180 Rio de Janeiro --
RJ, Brazil} 

\author{G.I. Gomero}
\address{Instituto de F\'{\i}sica Te\'orica, 
Universidade Estadual Paulista, \\ 
Rua Pamplona 145, \ \ 
01405-900 S\~ao Paulo -- SP, Brazil} 

\maketitle

It is well known that general relativity is a local metrical 
theory and therefore the corresponding field equations do not 
fix the global topology of spacetime.%
\footnote{By topology of the universe we mean the topology of 
the space-like section $M$.} 
This freedom has fuelled a great deal of interest in the possibility 
that the universe may possess spatial sections with non-trivial 
topology (see for example Refs. \refcite{CosmicTop} and
\refcite{Revs}). 

An immediate observational consequence of a nontrivial topology 
(multiple-connectedness) of the 3-space $M$ is that the sky 
may show multiple images of radiating sources. We are 
assuming here and in what follow that the universe has a 
detectable (nontrivial) topology (for details on this point 
see Ref. \refcite{Detect}).

However, the direct identification of multiple images is a formidable 
observational task to carry out because it involves a number of 
problems. This has motivated the development of methods 
in which the cosmic images are statistically treated in the search 
for a sign of a possible nontrivial topology of the universe.
In a universe with nontrivial topology the $3$--D positions 
of the multiple images are correlated, and these correlations can be 
couched in terms of pair-separation correlations.
The first statistical method (cosmic crystallography) looks for these 
correlations by using pair separations histograms (PSH).\cite{LeLaLu1996}
But the only significant (measurable) sign of a nontrivial 
topology in PSH's was shown to be spikes, and they can be used 
merely to reveal a possible nontrivial topology 
of universes that admit Clifford translations (for details see, 
e.g. Ref. \refcite{spikes}). 

The determination of the positions of cosmic sources, however, 
involves inevitable uncertainties, some of which have been discussed 
by Lehoucq \emph{et al.\/}.\cite{LeUzLu2000} 
Here we briefly report our results concerning the 
sensitivity of the topological spikes in the presence of 
the uncertainties in the positions of sources,
which arise from uncertainties in the values of the density 
parameters.

For brevity, we shall consider only flat universes
($\Omega_{m0}+\Omega_{\Lambda 0}=1$), but a similar analysis
can be carried out for spherical universes, with
qualitatively the same results. \cite{BernuiGomeroMotaReboucas} 
The redshift-distance relation for the flat case reads
\begin{equation} \label{red-dis}
r(z, \Omega_{m0})=\frac{c}{H_0} \int_1^{1+z}\!
\left[(x^3-1) \Omega_{m0}+ 1 \right]^{-1/2} dx\;.
\end{equation}
Consider now two cosmic sources at distances $r_1$ and $r_2$ from 
the observer $O$, with their lines of sight forming an angle 
$\theta$. The law of cosines gives the pair-separation $s=d^2$ of 
these objects 
\begin{equation}
\label{FlatCosLaw}
s = r_1^2 + r_2^2 - 2\,r_1r_2 \cos \theta \qquad (r_1 \leq
r_2) \;. 
\end{equation}
The uncertainty in $s$ which arises from the uncertainties in 
$\Omega_{m0}$ is
\begin{equation}
\Delta s = \frac{ds}{d \Omega_{m0}}\Delta \Omega_{m0}= 
2 \left[ (r_1 - r_2 \cos \theta) \frac{\partial r_1}
{\partial \Omega_{m0}}+ (r_2 - r_1 \cos \theta) 
\frac{\partial r_2}{\partial \Omega_{m0}} \right]
\Delta \Omega_{m0} \;.
\end{equation}

\begin{figure}[tbh]
\centerline{\def\epsfsize#1#2{0.4#1}\epsffile{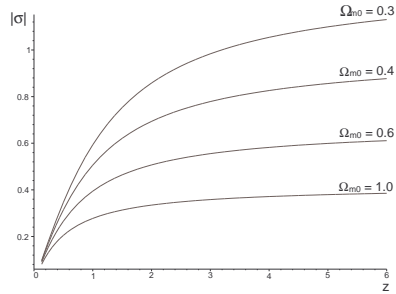}}
\caption{\small Curves of $|\sigma|$ versus $z$ for different 
$\Omega_{m0}$.}
\label{SigmaFig}
\end{figure}

For a fixed pair separation $s$ the uncertainties $\Delta s^{(1)}$
and $\Delta s^{(2)}$ at, respectively, $r_1$ and $r_2$ are such that
$|\Delta s^{(2)}| \geq | \Delta s^{(1)}|$.
For $r_1=r_2=r$ the relative error $\sigma$ is 
\begin{equation} \label{sigma}
\frac{\Delta s}{s} = 2\, \sigma \,\Delta\Omega_{m0}
\qquad \mbox{with} \qquad
\sigma(z,\Omega_{m0}) = \frac{\partial \ln r }{\partial \Omega_{m0}} \; .
\end{equation}
Figure \ref{SigmaFig} shows the behaviour of the relative error for
different values of $\Omega_{m0}$.
It makes apparent that the error $|\sigma|$ grows with $z$ and 
is lower in universes with smaller $\Omega_{\Lambda 0}$. 
The smallest values for the error $|\sigma|$ correspond to is 
for Einstein de--Sitter model ($\Omega_{\Lambda 0}=0$.
These curves also suggest $|\sigma|$ is upper bounded, and in 
fact, it can be shown that 
$\lim_{z \to \infty} \frac{\partial \sigma}{\partial z} = 0$.

{}From (\ref{red-dis}) and (\ref{sigma}) it is clear
that the error $|\sigma|$ in the determination of the positions
grows with $r$. In practice (real world) equal separations $s$
of correlated pairs (used in PSH) change. This can have basically 
the following effects: (i) spread the pair separations enough to 
destroy the spikes; (ii) spread the pair separations and move the 
spikes, without destroying them. 
These possibilities depend on both the error and also on the bin 
size. A suitable compromise between these variables (error $|\sigma|$ 
and bin size $\delta s = 4\, s_{max}^2/m$) can, however, be found.%
\cite{BernuiGomeroMotaReboucas}

\begin{figure}[tbh]
\centerline{\def\epsfsize#1#2{0.6#1}\epsffile{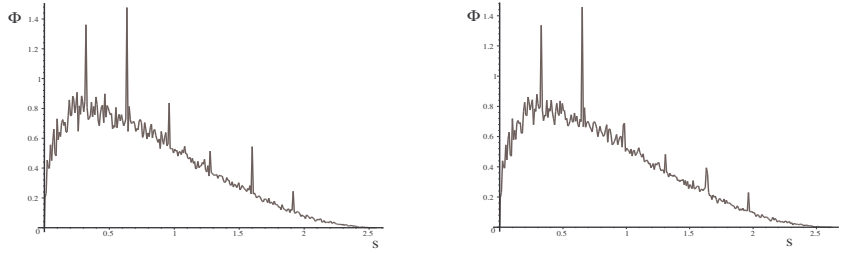}}
\caption{\small PSH's $\Phi(s)$ of a flat universe, whose 
corresponding topology is a cubic torus, for `exact' (left) and 
`approximate' values (right) values of density parameters.}
\label{PSH}
\end{figure}

Figure \ref{PSH} exhibits two PSH  cubic torus of side 
$L= 2\,\sqrt{2}/5$ for `exact' values of density parameters 
$\Omega_{m0}^{(e)}=0.3,\, \Omega_{\Lambda0}^{(e)}=0.7$ (left) and 
for `approximate' values $\,\Omega_{m0}=0.28,\,\Omega_{\Lambda0}=0.72$ 
(right). The bin size $\delta s$ is fixed by $s_{max}=0.8$ and $m=300$.
A close inspection of this figures makes clear that, for a fixed suitable
bin size, when one considers uncertainties in the density parameters: 
(i) the spikes are preserved but the amplitudes of the spikes decrease 
(this effect is clearer for large values of the pair-separation, for
which the errors are larger); 
(ii) the spikes are spread (consistent with the decreasing of their
amplitudes) and moves to the right.
In brief, the uncertainties in the density parameters break the 
degeneracies in pair separations due to translations, and put limits 
on the bin size $\delta s$ of PSH's: it has to be chosen large 
enough for not to resolve pair separations differences that arises 
from these uncertainties, but small enough not to include 
uncorrelated pair separations. However, it is always possible to 
conveniently choose the bin size so that the spikes are robust 
with respect to uncertainties in the density 
parameter.\cite{BernuiGomeroMotaReboucas}

To close this work, we mention that in the collected correlated pairs (CCP) method\cite{UzLeLu} an indicator of a detectable nontrivial topology of the 
spatial section $M$ of the universe is the CCP index 
\begin{equation}
\mathcal{R_{\epsilon}} = \frac{\mathcal{N_{\epsilon}}}{P-1} \; ,
\end{equation}
where $\mathcal{N_{\epsilon}} = Card\{i \,:\, \Delta_i \leq
\epsilon\}$, and $\epsilon > 0$ is a parameter that deal with 
the uncertainties in the determination of the pairs separations.
As in the PSH case, the CCP method also rely on the knowledge of 
the $3$-D positions of the cosmic sources, which again involve 
uncertainties. Therefore, the sensitivity of the CCP index in the 
presence of the uncertainties in the positions of the sources, 
which arise from uncertainties the density parameters, can similarly 
be made and is under development.\cite{BernuiGomeroMotaReboucas}


{\bf Acknowledgments:} We thank CNPq and FAPESP for financial support.

\end{document}